\documentclass[preprint]{revtex4}  
\usepackage{amsfonts}
\usepackage{amsmath}
\usepackage{cancel}
\usepackage{graphicx}
\usepackage{mathcomp}
\usepackage{color}

\newcommand{\sca}{\!\cdot\!}

\newcommand{\grad}{\nabla}
\newcommand{\dvg}{\nabla\!\cdot\!}

\begin{document}
\title{Active plasma resonance spectroscopy: \\ A functional analytic description}
\author{M. Lapke}\author{J. Oberrath}\author{T. Mussenbrock}\author{R.\,P. Brinkmann}
\affiliation{Institute for Theoretical Electrical Engineering, Ruhr University Bochum, Center for Plasma Science and Technology, D-44780 Bochum, Germany}
\date{\today}

\begin{abstract}
The term ``Active Plasma Resonance Spectroscopy'' denotes a class of diagnostic methods which employ the 
ability of plasmas to resonate on or near the plasma frequency. The basic idea dates back to the early 
days of discharge physics:  A signal in the GHz range is coupled to the plasma via an electrical probe; the spectral response is recorded, and then evaluated with a mathematical model to obtain information on the electron density and other plasma parameters. In recent years, the concept has found renewed interest as a basis of industry 
compatible plasma diagnostics. This paper analyzes the diagnostics technique in terms of a general description based on functional analytic (or Hilbert Space) methods which hold for arbitrary probe geometries. 
It is shown that the response function of the plasma-probe system can be expressed as a matrix element of the 
resolvent of an appropriately defined dynamical operator. A specialization of the formalism to a
symmetric probe design is given, as well as an interpretation in terms of a lumped circuit model consisting
of series resonance circuits. We present ideas for an optimized probe design based on geometric and electrical symmetry.
\end{abstract}

\maketitle

\section{Introduction}

Of the many diagnostic techniques available or proposed for
low temperature plasmas, only a few are compatible with industrial requirements. 
A diagnostic tool which is useful for supervision and/or control of
technological plasma processes must be i) robust and stable, ii) 
insensitive against perturbation by the process, iii) itself not 
perturbing to the process, iv) clearly and easily interpretable
without the need of calibration, v) compliant with the requirements 
of process integration, and, last not least, vi) economical in terms
of investment, footprint, and maintenance.

In this paper we focus on a very promising approach to industry
compatible plasma diagnostics, the so-called plasma resonance spectroscopy. 
This approach attempts to exploit the natural ability of plasmas to
resonate on or near the electron plasma frequency
$\omega_{\rm pe}=\sqrt{e^2 n_{\rm e}/\epsilon_0 m_{\rm e}}$ for
diagnostics purposes. The idea to use these resonances to determine plasma parameters dates back to the early days of discharge physics
\cite{tonks1929} and has found many different realizations since then
\cite{takayama1960, stenzel1960, cohen71, klick1996, sugai1999,
nakamura2003, kim2003, piejak2004, franz2005, booth2005, scharwitz2008,
scharwitz2009, li2010, wang2011}. We follow the classification of \cite{lapke11}: {\it Active plasma resonance
spectroscopy} couples a suitable RF signal into the plasma and records
the response \cite{takayama1960, stenzel1960, cohen71, sugai1999,
nakamura2003, kim2003, piejak2004, booth2005, scharwitz2008, scharwitz2009,
li2010, wang2011}. {\it Passive plasma resonance spectroscopy} merely observes
the pre-existing excitation of the plasma by other sources (typically,
the RF power input) \cite{klick1996, franz2005}. {\it Electromagnetic concepts}
are based on the interaction of the plasma with the full electromagnetic
field, they operate above the plasma frequency
and observe the plasma-induced shift of transmission line or cavity
resonances which are already present in vacuum \cite{stenzel1960, kim2003,
piejak2004, wang2011}. {\it Electrostatic concepts}, on the other hand, result
from the interaction of the plasma with the electric field alone; they
utilize resonances  below $\omega_{\rm pe}$ which are not present in
vacuum (standing surfaces waves) and deduce the plasma parameters from
the absolute value of the frequency \cite{takayama1960, cohen71, sugai1999,
nakamura2003, booth2005, scharwitz2008, scharwitz2009, li2010}.
Finally, {\it one-port concepts} utilize probes with only one electrical
input/output port \cite{takayama1960, cohen71, klick1996, sugai1999,
nakamura2003, piejak2004, franz2005, scharwitz2008, scharwitz2009, li2010,
wang2011}, while {\it two-port concepts} use a probe configuration with two ports
or even two separate probes \cite{stenzel1960, kim2003, piejak2004,
booth2005}. (Multi-port concepts would also be conceivable.)

Many groups have dealt with simulation and modeling of specialized aspects
of such diagnostic devices to achieve a better understanding of the
resonance behavior \cite{booth2005, scharwitz2008, xu2009, xu2010,
bin2010, liang2011}. In contrast, the purpose of this manuscript is to
study the whole class of active, electrostatic methods in order to
gain a deeper physical understanding of the resonance behavior. Therefore, we use
a general description based on functional analytic methods which allows for an analysis without particular assumptions on the probe geometries or the homogeneity of the plasma.

\section{A model of an electrostatic probe in arbitrary geometry}

\begin{figure}[h!]
\includegraphics[width= 0.8\columnwidth]{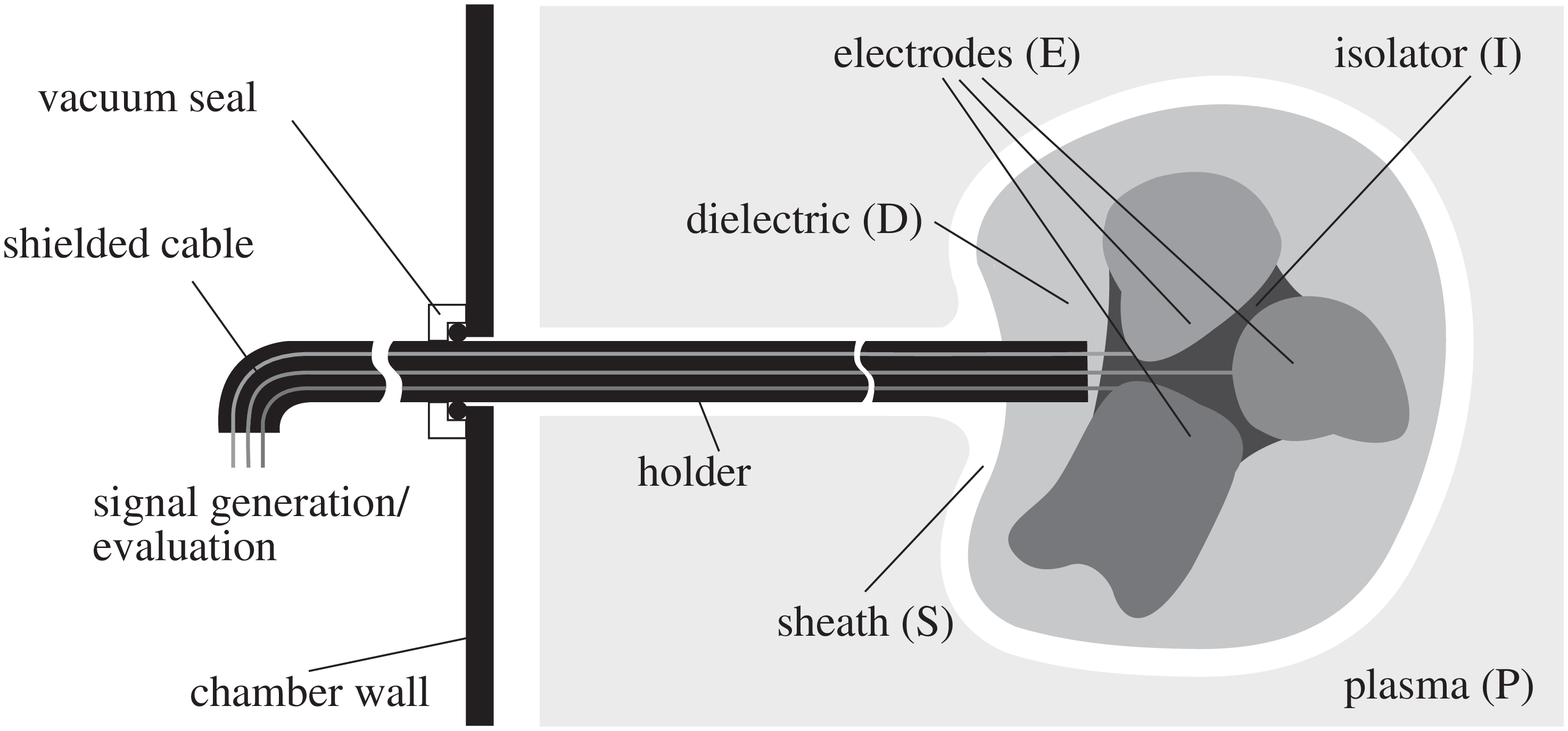}
\caption{Illustration of the abstract model for a N-electrode system. The electrodes are shielded to each other and the plasma by some dielectric medium. The whole probe, surrounded by a plasma sheath is immersed in a plasma volume.}\label{abstractmodel}
\end{figure}

We represent the plasma chamber (see figure \ref{abstractmodel}) as a simply connected, spatially
bounded domain $V$, most of which is plasma (a simply connected subdomain $P$). 
Other subdomains of $V$ are the plasma boundary sheath $S$, which shields the plasma from all material objects, 
and possibly dielectric domains $D$. The boundary $\partial V$ of the domain $V$ is either grounded ($G$) \linebreak
or ideally insulating ($I$) with vanishing conductivity and permittivity.
 
Into this idealized plasma chamber, a diagnostic device is immersed. To render our
con\-siderations general, we assume that it is a probe of arbitrary shape, containing an arbitrary number of powered electrodes. 
The electrodes surfaces $E_k$, $k=1\dots N$ are also seen as part of the domain boundary $\partial V$,
of course they are insulated from each other and from ground. A possible dielectric shielding of the probe is represented
as a part of the subdomain $D$ within the plasma chamber $V$. Fig.~1 illustrates the assumed geometry.

Within the subdomain $P$, the plasma is described by the
cold plasma model, a~set of two coupled linear equations for the RF
charge density $\rho$ and the RF current density $\vec{j}$. \linebreak
The first expresses the conservation of charge, the second describes
the acceleration of the electron fluid by the RF electric field
$\vec{E}=-\grad\Phi$ and the momentum loss due to electron-neutral
collisions (measured by its rate $\nu$), 
\begin{eqnarray}
&&\frac{\partial\rho}{\partial t} + \dvg \vec j = 0,\label{dyna} \\
  &&\frac{\partial\vec j}{\partial t} = -\epsilon_0 
  \omega_{\rm pe}^2 \grad \Phi
  - \nu{\vec j}.
  \label{dynb}
\end{eqnarray}
The electric potential $\Phi$ is not only defined in the plasma
domain $P$, but also in the sheath $S$\linebreak and in the dielectric $D$. 
RF charges are present in the plasma $P$ and -- as surface charges -- \linebreak 
on the plasma boundary $\partial P$. The sheath, however, is electron-free and hence RF charge free.
\linebreak Consequently, Poisson's equation 
gives the RF potential as  
\begin{equation}
-\dvg \left(\epsilon_0\epsilon \grad\Phi \right) = 
\left\{  
\begin{array}{ll}
\rho & \hbox{in the plasma $P$ and on $\partial P$, }\\[0.5ex]
0 & \hbox{in dielectric domains $D$ and the sheath $S$.}\\ 
\end{array}\right. \label{Poisson}
\end{equation}
The permittivity is given by
\begin{equation}
\epsilon= \epsilon{(\vec{r})}= \left\{  
\begin{array}{ll}
1 & \hbox{in the plasma $P$ and the sheath $S$, }\\
\epsilon_r(\vec r) & \hbox{in dielectric domains $D$.}
\end{array}\right.
\end{equation}

The electrodes $E_k$ are driven by RF voltages
$V_k$, with $k=1\dots N$. On the grounded parts of the wall
or the probes the potential vanishes. It is convenient to treat
these sections as a further electrode $E_0$ with an applied
voltage $V_0=0$. Taking also into account the insulating surfaces $I$,
the boundary conditions for the potential are of mixed type: 
\begin{equation}
	 \begin{array}{l@{\;\;\;\;\;}l}
	    \text{Electrodes $E_k$:}  & \Phi\vert_{E_k} = 
	    V_k, \;\;\;\;\; k=0\ldots N \\
	    \text{Insulators $I$:} & \vec{n}
	    \sca\grad\Phi\vert_{I} = 0.  
	 \end{array}
	 \label{BoundaryConditions}
\end{equation}

The currents $I_k$ carried by the electrodes are the
measured signals of the diagnostic. As~the electrodes 
are shielded from the plasma, at least by the
sheath, the currents are carried by displacement alone. Counting
them positive when they flow from the electrode to the plasma (in the direction of the negative surface normal), 
one obtains
\begin{equation}
I_k = \int_{E_k} \epsilon_0 \epsilon_r \frac{\partial\grad\!\Phi}{\partial t} \,\sca\,d^2\vec{r}.\label{current}
\end{equation}

Of course, there is a standard approach to solve these equations: Applying the Fourier transform to the equation
of motion leads to the RF version of Ohms law, an expression of the total current in terms of the field. 
The model reduces to an elliptic equation in the domain $V$,
\begin{equation}
 \dvg \left(\epsilon_0\epsilon^{\rm (eff)} \grad \Phi\right) 
 = 0,\label{NumericalPoisson}
\end{equation}
with the effective permittivity $\epsilon^{\rm (eff)}$ given by
\begin{equation}
\epsilon^{\rm (eff)} = 
\left\{  
\begin{array}{ll}
1    -\frac{\displaystyle \omega_{\rm pe}^2}{\displaystyle 
\omega(\omega -i\nu)} & \hbox{in the plasma $P$, }\\[0.5ex]
1 & \hbox{in the sheath $S$,}\\[.5ex] 
\epsilon_r(\vec r) & \hbox{in dielectric domains $D$.}
\end{array}\right.
\end{equation}
Together with the Fourier transformed boundary conditions \eqref{BoundaryConditions}, the elliptic equation can easily 
be solved numerically. Its linearity allows to write the solution $\Phi(\vec{r})$ as a superposition of 
$N$ fundamental solutions, 
\begin{equation}
\Phi(\vec{r}) = \sum_{l=1}^N V_l \Phi_l(\vec{r}).
\end{equation}
each of which obeys equation \eqref{NumericalPoisson} and the boundary conditions
($\delta_{kl}$ is Kronecker's delta)
\begin{eqnarray}
 \begin{array}{l@{\;\;\;\;\;}l}
	    \text{Electrode $E_k$:}  & \Phi_l\vert_{E_k} = 
	    \delta_{kl}, \;\;\;\;\; l=0 \ldots N , \\
	    \text{Insulators:} & \vec{n}\sca\grad\Phi_l
	    \vert_{\partial I} = 0  .
	 \end{array}
	 \label{NumericalBoundaryConditions}
\end{eqnarray}

With the help of the fundamental solutions $\Phi_l$, a response matrix $Y_{kl}$ can be defined as the
flux of the Fourier transformed displacement current
through electrode $E_k$,
\begin{equation}
Y_{kl}=\int_{E_k} i\omega\epsilon_0\epsilon^{\rm (eff)} \grad\Phi_l\,\sca d^2\vec r\,.
\end{equation} 
The Fourier transformed RF current $I_k$ as the probe signal can then be calculated as
\begin{equation}
	I_k =\sum_{l=1}^N Y_{kl} V_l,\label{NumericalCurrent}.
\end{equation}

This approach has been successfully employed for numerical investigations of the plasma absorption probe
and the multipole resonance probe in \cite{lapke08} and
\cite{lapke07}. It relies, however, on numerical calculations
and gives only a limited insight into physics of the probe 
response: Although the mathematical structure of 
\eqref{NumericalPoisson} to \eqref{NumericalCurrent} is
quite transparent, it is not possible to separate the
plasma contributions from the influences of the geometry
from the numerical solution.

\section{Current and energy balance equations}

In order to obtain a deeper understanding of the probe model behavior, we focus on the balance equations of current and energy and continue with a general formalism. To make this section more readable some derivations are shifted to the appendix. We
split the electric potential into two parts,
one of which is the vacuum potential which can be written
as a linear superposition of characteristic functions,
\begin{equation} 
\Phi(\vec{r}) = \phi(\vec{r}) + \Phi^{\rm (vac)}(\vec{r})
 = \phi(\vec{r})+ \sum_{l=0}^N V_l \Psi_l(\vec{r})\,.
 \label{charfunc}
\end{equation}
The characteristic functions $\Psi_l$, with $l=0\dots N$,
depend only on the geometry of the chamber and the geometry
of the probe. They are independent of the plasma and its
spatial distribution. To calculate the $\Psi_l$, the
related Laplace equation
has to be solved, i.e. 
\begin{equation}
\nabla \sca \left(\epsilon_0 \epsilon_r \nabla \Psi_{l} \right) =0\ .
\end{equation} 
The boundary conditions are given by  
\begin{eqnarray}
 \begin{array}{l@{\;\;\;\;\;}l}
	    \text{Electrode $E_k$:}  & \Psi_l\vert_{E_k} = 
	    \delta_{kl}, \;\;\;\;\; l=0 \ldots N ,\\
	    \text{Insulators:} & \vec{n}\sca\grad\Psi_l
	    \vert_{\partial I} = 0  .
	 \end{array}
	 \label{VacuumBoundaryConditions}
\end{eqnarray}

The first part of the electric potential defined in \eqref{charfunc}, 
referred as the inner potential $\phi$, is given by Poisson's equation
in the plasma and the sheath, or dielectric zones, and satisfies homogeneous 
boundary conditions. One obtains the boundary value problem
\begin{eqnarray}
&&-\dvg \left(\epsilon_0\epsilon_r \grad\phi \right) = 
\left\{  
\begin{array}{ll}
\rho & \hbox{in the plasma $P$ and on $\partial P$, }\\[0.5ex]
0 & \hbox{in dielectric domains $D$ and the sheath $S$,}\\[.5ex] 
\end{array}\right. \label{InnerPoisson} \\
&&	 \begin{array}{l@{\;\;\;\;\;}l}
	    \text{Electrode $E_k$:}  & \phi\vert_{E_k} = 0, \;\;\;\;\; k=0\ldots N, \\
	    \text{Insulator:} & \vec{n}\sca\grad\phi\vert_{\partial I} = 0.  
	 \end{array}
	 \label{InnerBoundaryConditions}
\end{eqnarray}
With the knowledge of the two distinct parts of the potential, we are able 
to reformulate dynamical equations for the charge density and the current
density. From \eqref{dyna} and \eqref{dynb} we ultimately arrive at
\begin{eqnarray}
&&\frac{\partial\rho}{\partial t} = -\dvg \vec j, \label{Dyn1} \\
  &&\frac{\partial\vec j}{\partial t} = -\epsilon_0 \omega_{\rm pe}^2 \grad \phi- \nu{\vec j}
  -\epsilon_0 \omega_{\rm pe}^2\sum_{l=0}^N V_l \grad \Psi_l.\label{Dyn2}
\end{eqnarray}

Similarly to the potential, the current $I_k$ carried by the electrode
$E_k$ can also be split into two parts, $I_k(t) = i_k(t) + I_k^{\rm (vac)}(t)$.
The first term is referred to as the inner current, the second is the vacuum part.
The vacuum part is nothing but the displacement current related to 
the vacuum field $\Phi^{\rm (vac)}$. It is shown in the appendix that
it has a purely capacitive character. One obtains
\begin{equation}
I_k^{\rm (vac)} = -\sum_{l=0}^N C_{kl}\frac{\partial V_l}{\partial t}. 
\end{equation}
The coefficients $C_{kl}$ are the capacitance coefficients defined by
\begin{equation}
	C_{kl} =   \int_{\partial V}\Psi_k \epsilon_0\epsilon_r\grad\Psi_l \sca d^2\vec{r}.\label{capcoeff}
\end{equation}

It is also shown in the appendix that the vacuum current satisfies 
Kirchhoff's law of current conservation, i.e., the sum of the currents 
over all electrodes (including grounded electrodes) is equal to zero,
\begin{equation}
\sum_{k=0}^N I_k^{\rm(vac)} =  \sum_{k=0}^N \sum_{l=0}^N  C_{kl}
\frac{\partial V_l}{\partial t}= 0.
\end{equation}
With the help of the characteristic functions introduced in the previous
section, the inner part of the current $i_k$ can be written as a flux integral over the electrode $E_k$ 
\begin{eqnarray}
i_k = - \int_V \grad\Psi_k \sca \vec{j}\, d^3r. \label{innercurrent}
\end{eqnarray}
(Again the explicit derivation can be found in the appendix.) The inner part of the current obeys Kirchhoff's law of
current conservation, i.e., their sum over all electrodes (including
ground electrodes) is equal to zero (see also appendix).

We now establish energy balance equations for the system. Owing to
the fact that the coupling of the vacuum field and the inner potential
vanishes (see appendix), we can split the electrical energy of the system 
into a vacuum part $E^{\rm (vac)}$ and an inner part $ \varepsilon^{\rm (el)}$ given
by
\begin{eqnarray}
\varepsilon^{\rm (el)}&=& \int_V \frac{1}{2}  \epsilon_0\epsilon_r \grad\phi\sca\grad\phi\,d^3r
\end{eqnarray}
and
 \begin{eqnarray}
E^{\rm (\rm vac)}= \frac{1}{2}\sum_{k=0}^N \sum_{l=0}^N  C_{kl} V_k V_l\,.
\end{eqnarray}
The vacuum part can be expressed in terms of the previously defined capacitance
coefficients \eqref{capcoeff}. Its derivative with respect to time simply
describes the charging and decharging of the linear multiport capacitor $C_{kl}$,
\begin{equation}
	 \frac{dE^{\rm (vac)}}{dt} = \sum_{k=0}^N \sum_{l=0}^N  C_{kl} V_k
	 \frac{\partial V_l}{\partial t}
	           = \sum_{l=0}^N V_l I^{\rm( vac )}.
\end{equation}
The inner energy $\varepsilon$ is the sum of the inner electrical energy 
$ \epsilon^{\rm (el)}$ and the kinetic energy due to the motion of electrons 
 \begin{equation}
	\varepsilon = \varepsilon^{\rm (el)}+\varepsilon^{\rm (kin)}=  \int_V \frac{1}{2}  \epsilon_0\epsilon_r \grad\phi\sca\grad\phi\,d^3r +
	          \int_P  \frac{1}{2\epsilon_0\omega_{\rm pe}^2} \vec{j}^{\,2}\, d^3 r.\label{innerenergy}
\end{equation}
The temporal change in the inner energy is governed by the balance of the power input
and the Ohmic dissipation due to collisions
\begin{eqnarray}
	\frac{d\varepsilon}{dt} =
	\sum_{l=0}^N V_l i_l 
  - \int_P  \frac{\nu}{\epsilon_0\omega_{\rm pe}^2} \vec{j}^2 \, d^3 r. 
\end{eqnarray}

\section{A lumped element analogy}

\begin{figure}[h!]
\includegraphics[width= 0.8\columnwidth]{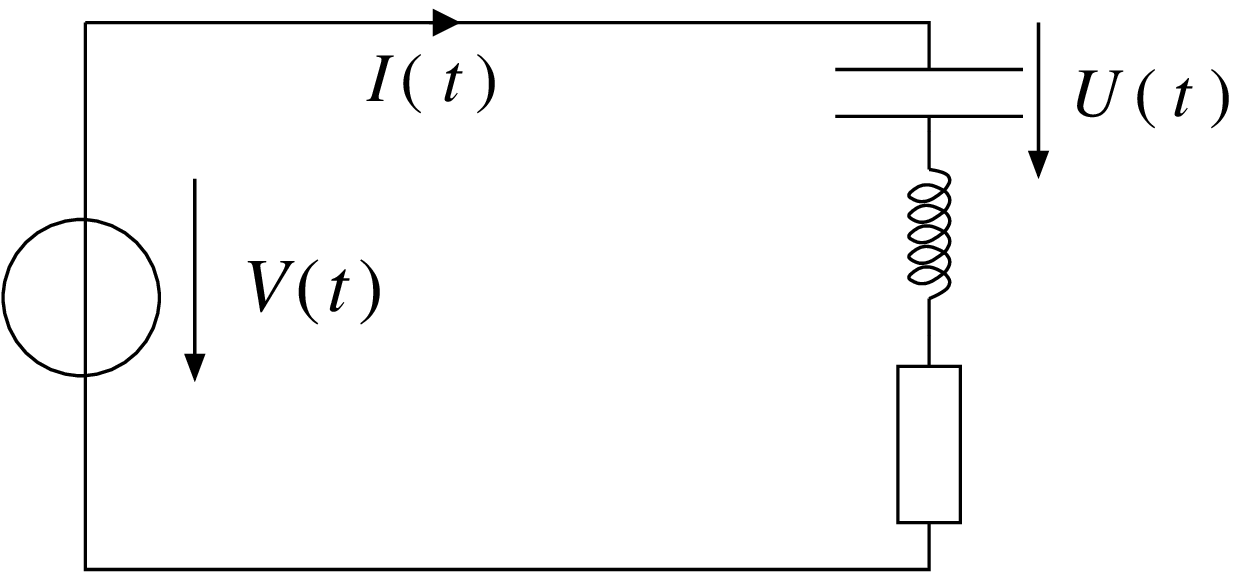}
\caption{A lumped series resonance circuit driven by a voltage $V(t)$. This circuit is interpretable as an one-port concept of active resonance spectroscopy which does not couple to the outer ground.}\label{rcl}
\end{figure}

As a simple problem and to introduce an abstract mathematical formulation, we consider the response of a lumped series resonance circuit driven by a voltage $V(t)$ as shown in figure \ref{rcl}. One can interpret this scenario as a one-port concept which does not couple to the outer ground. We assume the values of the lumped circuit elements as given.
(To evaluate real plasma diagnostic concepts the values have of course to be
determined by means of a mathematical model.)

The exemplary model can be cast as a system of two coupled ordinary differential 
equations of first order for the capacitor voltage $U$ and the current $I$,
\begin{eqnarray}
\frac{dU}{dt} &=&  \frac{I}{C},\\[0.5ex]
\frac{dI}{dt} &=& -\frac{U}{L} - \frac{RI}{L} + \frac{1}{L} V(t). 
\end{eqnarray}
To solve the system the Fourier transform can be used and one obtains its admittance that is interpretable as the system response of the one-port probe 
\begin{equation}
  Y(\omega)= \frac{I}{V} =\displaystyle \frac{\displaystyle 1}{\displaystyle R + i\omega L + \frac{\displaystyle 1}{\displaystyle i\omega C}}.	\label{responseRCL}
\end{equation}

The same result can be derived by an abstract mathematical formulation of the system. To introduce this formulation we refer to the ordered pair of $U$ and $I$ as the state of the system,
denoted by the state vector
\begin{equation}
	\vert z \rangle \equiv \vert U,I \rangle = \left(\!\begin{array}{c}U \\ I\end{array}\!\right).
\end{equation}
Similarly, we introduce the vector of the excitation of the system
\begin{equation}
	\vert e \rangle =  \left(\!\begin{array}{c}0 \\ \displaystyle\frac{1}{L}\end{array}\!\right).
\end{equation}
The operators of the non-dissipative and the dissipative dynamics are given
by
\begin{equation}
T_C = \left(\begin{array}{cc}0 & \displaystyle \frac{1}{C} \\ -\displaystyle\frac{1}{L} & 0\end{array}\right), \qquad
T_D = \left(\begin{array}{cc}0 & 0 \\ 0 &- \displaystyle \frac{R}{L}\end{array}\right).
\end{equation}
With the help of these definitions, the behavior of the lumped element circuit
can be expressed in terms of a compact dynamic equation,
\begin{equation}
	\frac{d\vert z\rangle}{dt} = T_{\rm C} \vert z\rangle + T_{\rm D} \vert z\rangle + 
	 V(t)\vert e\rangle\label{DynRCL}.
\end{equation}

Since the set of all states $\vert z\rangle$ forms a vector space $Z$ it is
advantageous to employ as the underlying scalar field the set of complex
numbers $\mathbb{C}$. To turn $Z$ into a Hilbert space, the appropriate
scalar product has to be defined. Due to the fact that the dimension of the
vectors $\vert z\rangle$ is inhomogeneous, the Euclidian product is not
available. However, a physically reasonable quadratic form exists, i.e.,
the energy $\varepsilon =\left(C U^2 + LI^2\right)/2$, which is stored in the
capacitance and the inductance of the resonant circuit. This motivates
the definition of a scalar product of two state vectors $\vert\tilde z\rangle$
and $\vert z\rangle$ as
\begin{equation}
	\langle \tilde z\vert z\rangle \equiv 	\langle \tilde U,\tilde I\vert U,I\rangle :=
	C {\tilde U}^*U + L {\tilde I}^*I.\label{scalarRCL}
\end{equation}
The energy is a non-negative real quantity even for complex states,
\begin{equation}
 \varepsilon = \frac{1}{2} \left\|z\right\|^2 = \frac{1}{2}\langle z\vert z\rangle. \label{energyrcl}
\end{equation}
The defined scalar product is compatible with the dynamical equations.
It can be shown that the operator $T_C$ is anti-Hermitian with respect
to the scalar product,
\begin{equation}
	\langle  T_C \tilde z\vert z\rangle  = - \langle \tilde z\vert T_C  z\rangle, 
\end{equation}
while the dissipative operator $T_D$ is Hermitian and negative definite,
\begin{eqnarray}
\langle  T_D \tilde z\vert z\rangle&=& \langle \tilde z\vert T_D\vert z\rangle, \\
\langle \tilde z\vert T_C\vert z\rangle &\le& 0.
\end{eqnarray}
Furthermore, it can be shown that the observable response of the system, i.e.,
the current $I$, is the scalar product of the excitation vector and the 
state vector $\vert z\rangle$. Thus, the observation vector is equal to
the excitation vector,
\begin{equation}
I= \langle 0,{\frac{\displaystyle 1}{\displaystyle L}}\vert U,I\rangle = \langle e \vert z\rangle .
\end{equation}
The energy balance equation \eqref{energyrcl} can be transformed to the energy balance equation of a series resonance circuit
\begin{equation}
	\frac{d}{dt}
	\left(
	\frac{1}{2}C U^2 +  \frac{1}{2} LI^2 \right)= - R I^2 + I \, V(t).
\end{equation}

To analyze the response of the system we apply the Fourier transform to 
\eqref{DynRCL}. The current is then given by
\begin{equation}
I = \langle e	\vert z\rangle  
	=\langle e \vert \frac{1}{\displaystyle i\omega - T_{\rm C} - T_{\rm D} }  \vert e\rangle V(\omega)
	= Y(\omega) V.\label{responseRCLabstract}
\end{equation}
The response function $Y(\omega)$ of the system can be directly specified by evaluating the matrix products. It should be noted, that two extra matrices from the scalar product \eqref{scalarRCL}
appear in the calculation. However, 
the response function can also be calculated by expanding the state vectors into the eigenfunctions of the operator. One ultimately obtains exactly the same expresion than in 
(\ref{responseRCL}). This result of the simple RCL-series circuit allows the physical interpretation of the abstract solution that is derived in a section further down.

\section{Functional analytic formulation of the probe response}

We are now equipped with all the material necessary for a functional
analytic formulation of the system of dynamical equations \eqref{Dyn1} and \eqref{Dyn2}. We interpret the charge density  $\rho(\vec{r})$ and the current density $\vec{j}(\vec{r})$ defined on the plasma domain $P$ as the variables of the system state
$\vert z\rangle$. (It is advantageous to allow for complex states.) In Dirac's bra-ket
notation we define an appropriate function space as
\begin{equation}
{\mathcal H}=\left\{ \vert z\rangle \equiv \vert \rho,\vec{j}\rangle\equiv
(\rho,\vec{j})^{\rm T} \right\} .
\end{equation}
For two system states
$\vert z\rangle \equiv \vert \rho,\vec{j}\rangle$ and
$\vert z^\prime\rangle \equiv \vert\rho^\prime,\vec{j}^\prime\rangle$
we define a scalar product $\left\langle  z|z^\prime  \right\rangle$
which is motivated by the inner energy functional \eqref{innerenergy}.
The following Hermitian form is compatible with the dynamical equations
and satisfies all requirements of an inner product, namely i) conjugate
symmetry, ii) sesquilinearity, and iii) positive definiteness. 
\begin{gather}
 \left\langle  z|z^\prime  \right\rangle =
 \int_V \epsilon_0 \,\epsilon_r   \nabla\phi\{\rho\}^* \sca \nabla \phi\{\rho^\prime\} \, d^3r +
 \int_P \frac{1}{\epsilon_0\omega_{\rm pe}^2}\vec{j}^* \sca \vec{j}^\prime \, d^3r\,.\label{scalar}
\end{gather}
The scalar product defines a corresponding norm $\left\|z\right\|$.
The norm-square of a state is equal to its inner energy, up to a factor of two
\begin{equation}
	\left\|z\right\|^2 = 
 \left\langle  z|z \right\rangle =
 \int_V \epsilon_0 \,\epsilon_r   \nabla\phi\{\rho\}^* \sca \nabla \phi\{\rho\} \, d^3r +
 \int_P \frac{1}{\epsilon_0\omega_{\rm pe}^2} \vec{j}^* \sca \vec{j}\, d^3r\label{scalar}
= 2\varepsilon\,.\end{equation}
Equipped with this scalar product, and suitably completed, the defined
function space of all possible states forms a Hilbert space $\mathcal{H}$. It should be noted that
the elements of $\mathcal{H}$ are not all regular functions. The charge
part may contain also  singular functions like surface charges. However, this is
acceptable as long as the norm $\left\|z\right\|$ is finite.

One important family of states $\vert e_k\rangle$ can be obtained
from the characteristic vacuum functions $\Psi_k$ which have only a current
part, their charge part and also the corresponding potential are equal to zero,
\begin{equation}
	\vert e_k\rangle = \vert 0,-\epsilon_0 \omega_{\rm pe}^2\grad\Psi_k\rangle. 
\end{equation}
The scalar product of  $\vert e_k\rangle$ with a system state $\vert z\rangle$
is an expression for the inner current $i_k$ which is carried by the
electrode $E_k$. Utilizing the definition of the scalar product \eqref{scalar}
one obtains
\begin{equation}
\langle e_k\vert z\rangle
=\int_V \grad\Psi_k \sca \vec{j}\, d^3r =i_k.	
\end{equation}

The dynamical equations \eqref{Dyn1} and \eqref{Dyn2} can be written in a compact form \cite{mussenbrock2007}. We define two dynamical
operators $T_C$ and $T_D$ which act on the Hilbert space $\mathcal{H}$.
The conservative operator $T_C$ contains the effects of the electron inertia and the electric field; it is defined by the action
\begin{equation}
T_C\vert \rho,\vec{j}\rangle =
\vert -\dvg \vec j,-\epsilon_0 \omega_{\rm pe}^2 \grad \phi\{\rho\}\rangle.	
\end{equation}
$T_C$ can be shown to be anti-Hermitian, i.e., $T_C=-T_C^\dagger$ (see appendix).
Thus the associated eigenvalues are purely imaginary and represent the frequency behavior.
The dissipative operator $T_D$ contains the effects of the collisions 
\begin{equation}
T_D\vert \rho,\vec{j}\rangle =
\vert 0,-\nu\vec{j} \rangle.	
\end{equation}
It can be shown that $T_D$ is Hermitian, i.e., $T_D=T_D^\dagger$
(see appendix). The eigenvalues are negative definite,
$\left\langle  z|T_D|z  \right\rangle\leq 0$. Furthermore, they are
real and contain information about damping of the system. With the
help of these definitions, the dynamical equations  \eqref{Dyn1}
and \eqref{Dyn2} of the system can be written as
\begin{equation}
	\frac{\partial\vert z\rangle}{\partial t}
	  = T_C \vert z\rangle + T_D \vert z\rangle + \sum_{l=0}^N V_l \vert e_l\rangle .		    \label{Dynabstract}
	  \end{equation}
Applying a Fourier transform to this abstract form and solving for $\vert z\rangle$,
the measured current can be obtained
\begin{equation}
i_k =\langle e_k \vert z\rangle
	  = \sum_{l=1}^N \langle e_k \vert\left(i\omega -T_C - T_D\right)^{-1}  \vert e_l\rangle V_l = \sum_{l=1}^N y_{kl} V_l \ .\label{abstractcurrent}
\end{equation}
Thus, the response function, can be written as
\begin{equation}
y_{kl}(\omega) = \langle e_k \vert\left(i\omega -T_C - T_D\right)^{-1} \vert e_l \rangle . \label{abstractresponse}
\end{equation}

The main result of this analysis is that the response function
$y_{kl}(\omega)$ is given by the matrix elements of the resolvent of the
dynamical operator equation evaluated for values on the imaginary axis.
The result can be ultimately interpreted using equivalent lumped element circuits as described above.

\section{Results for a one-port probe}

With this result and the insights gained from the simple application of the formulated 
theory, we are able to discuss the result of the abstract model, given
by equation \eqref{abstractcurrent} in terms of a lumped element circuit
model. For a one-port probe (we account for two electrodes plus ground) 
the current to one of the two powered electrodes is given as
\begin{equation}
i_k =\langle e_k \vert z\rangle
	  = \sum_{l=0}^2 \langle e_k \vert\left(i\omega -T_C - T_D\right)^{-1}  \vert e_l\rangle V_l = \sum_{l=0}^2 y_{kl} V_l .
\end{equation}
Comparing the response function $y_{kl}(\omega)$ \eqref{abstractresponse} with the one in \eqref{responseRCLabstract} reveals the same structure. It represents the coupling between
the powered electrodes directly and the coupling from each electrode
to ground. In this case, the dynamical operator $T=T_C+T_D$ is not a
matrix as in the simple example, but a functional operator. The 
current is able to flow on infinite different paths through the
plasma, each path can be expressed by a series resonance circuit.
Therefore, the response function represents an infinite number of
series resonance circuits connected in parallel.

\begin{figure}[h!]
\includegraphics[width= 0.8\columnwidth]{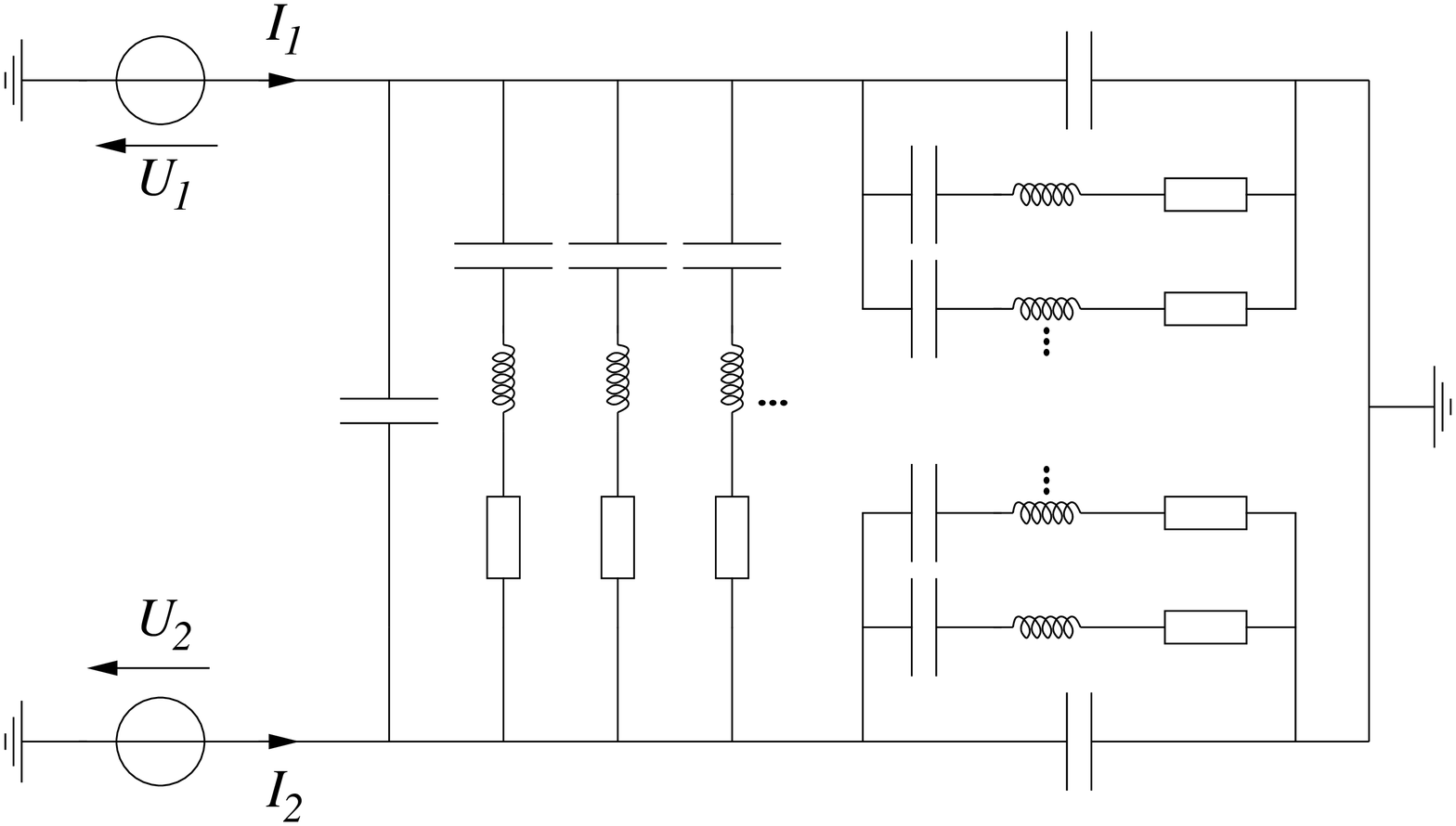}
\caption{Equivalent circuit for an one-port probe. It consists of a branch which represents the direct coupling between the two electrodes, and two additional branches which stand for the parasitic coupling of the electrodes to ground or ''infinity''. Each branch in itself may be complex, and the relation between the plasma and probe parameters and the values of the circuit elements is generally very complicated.}\label{abstractrcl}
\end{figure}

In summary, the complete equivalent circuit consists of a branch which 
represents the direct coupling between the two powered electrodes. Two additional branches mimic the parasitic coupling of the electrodes
to ground or ``infinity''. Each branch itself is complex, and the
relation between the plasma and probe parameters and the values
of the circuit elements are generally very complicated. This case is
depcited in figure \ref{abstractrcl}. For realizations with more than
one port the coupling gets even more complex. 

Under strictly electrical and geometrical symmetry the situation for a
one-port probe becomes simpler. It can be shown via a $\Delta$-Y
transform (circuit theory) that the coupling to ground or ``infinity'' vanishes.
As the field decays quickly with increasing distance from the probe, 
the measurements become actually local. The equivalent circuit becomes
more simple: an infinite number of series resonance circuits, each
of which representing a discrete resonance mode, are connected in 
parallel to a vacuum coupling.  

The multipole resonance probe proposed by the authors \cite{lapke08}
fulfills these requirements to a great extent -- it consists of two separated hemispherical electrodes covered by a dielectric shielding. Thus its geometry is approximately symmetric (with respect to rotations around the center of the probe) and its electrical behavior is completely symmetric (with respect to the mapping $\omega t \rightarrow \omega t + \pi$ and $\Phi\rightarrow -\Phi$. Additionally, the spherical geometry allows for a compact mathematical description of the resonance behavior and simple algebraic expression for its resonance frequencies.

\section{Summary / Conclusion / Outlook}
We have formulated a model of electrostatic active resonance spectroscopic methods in terms of a functional analytic approach. The abstract character of these methods allow for a generalized model where the geometrical and electrical details of the probe realization do not enter explicitly in the equations. We presented a general analytic solution for the system state that gives physical insight in the resonance behavior. The scalar product between this state and the observation vector was shown to be the inner current at one electrode. Part of the current is the inner admittance that is determined as a matrix element by the resolvent and the observation vector. This result can be interpreted in terms of a lumped element equivalent circuit. To guide the physical interpretation of the functional analytic approach this method was explicitly used to obtain the systems response of a idealized situation, i.e., that the equivalent circuit is described by a simple RCL-series circuit.

As an example we presented the equivalent circuit of an one-port probe. An analysis of this circuit allows for an optimization of the probe geometry. Due to a symmetrical geometry and a symmetrical applied voltage it is possible to transform the circuit in an equivalent circuit of one branch between the two electrodes. This branch is given by an infinite number of parallel series resonance circuits and each represents a resonance mode parallel to a vacuum coupling. 

Of course the presented theory can be extended in many ways. For instance, the influence of a magnetic field or kinetic effects may be taken into account. The magnetic field is negligible in the described theory as long as the skin depth is greater than the influence region of the probe and the gyration frequency is not in the range of the resonance frequency. A kinetic treatment of the electrons becomes important if the pressure is in the range of a few Pa and lower. First attempts to establish a kinetic theory based on functional analytic methods are in progress.

Non the less, the authors applied the formulation to the \emph{multipole resonance probe (MRP)} \cite{lapke08} which is the optimized concept. Its application area in a dielectric depositing plasma process is an advantage compared to a Langmuir probe. Measurements with the prototype of the MRP in a double ICP \cite{lapke11} have shown the feasibility of this concept and were in very good agreement with a Langmuir probe.

\appendix

\section{Derivation of the balance equations}

In the following, the balance equations and their relations will be formally derived. The calculations utilize the facts that the particle current vanishes at the electrodes due to the dielectric shielding, the total inner current is divergence-free, the $\Psi_k$ are the characteristic functions defined by eq. \eqref{charfunc} and employs the boundary conditions of $\Phi$ and $\phi$.\\

The vacuum current can be expressed via capacitance coefficients in the following form
\begin{equation}
I_k^{\rm (vac)} = -\int_{E_k} \epsilon_0\epsilon_r \frac{\partial \grad\Phi^{\rm (vac)}}{\partial t} \sca d^2\vec{r} 
    = -\int_{\partial V}\Psi_k \sum_{l=0}^N \frac{\partial V_l}{\partial t}\epsilon_0\epsilon_r\grad\Psi_l \sca d^2\vec{r} 
        = -\sum_{l=0}^N C_{kl}\frac{\partial V_l}{\partial t}. \label{appivac}
\end{equation}
It can also be shown, that Kirchhoff's law of current conservation is conserved.
\begin{eqnarray}
		\sum_{k=0}^N C_{kl} &=& \sum_{k=0}^N  \int_{\partial V}\Psi_k \epsilon_0\epsilon_r\grad\Psi_l \sca d^2\vec{r}
		=\sum_{k=0}^N\int_V \dvg (\Psi_k \epsilon_0\epsilon_r\grad \Psi_l)d^3r \nonumber\\
		&=&\sum_{k=0}^N \int_{V}\epsilon_0\epsilon_r\grad\Psi_k \sca\grad\Psi_l \,  d^3 r
		= \sum_{k=0}^N \int_{\partial V}\epsilon_0\epsilon_r\grad\Psi_k\Psi_l \sca \,d^2\vec{r} \nonumber\\
		&=&  \int_{\partial V}\epsilon_0\epsilon_r\grad\Psi_l \sca \,d^2\vec{r} = 
		\int_{V}\dvg\left(\epsilon_0\epsilon_r\grad\Psi_l \right) \,d^3 r =\,\label{appckl}
		 0.
\end{eqnarray}
The concise for the inner current can be derived as following
\begin{eqnarray}
i_k &=& - \int_{E_k} \epsilon_0\epsilon_r \frac{\partial \grad\phi}{\partial t} \sca d^2\vec{r}
=- \int_{\partial V} \Psi_k \left( \vec{j}-\epsilon_0\epsilon_r \frac{\partial \grad\phi}{\partial t}\right) \sca d^2\vec{r} \nonumber
    = -\int_V \grad\Psi_k \sca \left( \vec{j}-\epsilon_0\epsilon_r \frac{\partial \grad\phi}{\partial t}\right) d^3r\\
       &=&{-\int_V \grad\Psi_k \sca \vec{j} \, d^3r 
       +    \int_{\partial V} \grad\Psi_k \sca \left( \epsilon_0\epsilon_r\frac{\partial{\Phi}}{\partial t}\right) d^2\vec{r}}
= - \int_V \grad\Psi_k \sca \vec{j}\, d^3r. \label{appinnercurrent}
\end{eqnarray}
Also the inner current obeys Kirchhoff's law
\begin{equation}
	\sum_{k=0}^N i_k 
{=\sum_{k=0}^N  \int_{\partial V} \Psi_k \left( \vec{j}-\epsilon_0\epsilon_r \frac{\partial \grad\phi}{\partial t}\right) \sca d^2\vec{r}
=\int_{\partial V} \left( \vec{j}-\epsilon_0\epsilon_r \frac{\partial \grad\phi}{\partial t}\right) \sca d^2\vec{r} }
=0.\label{appinnercurrentzero}
\end{equation}
It can be shown, that the coupling of vaccum- and inner energy vanishes
\begin{eqnarray}
  E^{\rm (cpl)} &=&
	\int_V \epsilon_0\epsilon_r \grad\phi\,\sca\grad\Phi^{\rm(vac)} d^3r \nonumber \\
&=& 
{	\int_{\partial V} \phi\, \epsilon_0\epsilon_r \grad\Phi^{\rm(vac)}
	\sca d^2\vec{r} -
	\int_V \phi\,\dvg(\epsilon_0\epsilon_r \grad\Phi^{\rm(vac)})\, d^3r
	=0}. \label{enercpl}
\end{eqnarray}

For the inner energy, we obtain after some calculation the following relation which reflects the Ohmic dissipation due to the collisions:
\begin{eqnarray}
	\frac{d\varepsilon}{dt} &=& {\int_V  \epsilon_0\epsilon_r \grad\phi\sca\frac{\partial\grad\phi}{\partial t}\,d^3r +
	          \int_P  \frac{1}{\epsilon_0\omega_{\rm pe}^2} \vec{j}
	          \sca\frac{\partial\vec{j}}{\partial t}\, d^3 r} \nonumber\\[0.5ex]
	      &=&{  \int_V  \epsilon_0\epsilon_r \nonumber \grad\phi\sca\frac{\partial\grad\phi}{\partial t}\,d^3r +
	          \int_P  \frac{1}{\epsilon_0\omega_{\rm pe}^2} \vec{j}
	          \sca\left(-\epsilon_0 \omega_{\rm pe}^2 \grad \phi- \nu{\vec j} -\epsilon_0 \omega_{\rm pe}^2\sum_{l=0}^N V_l \grad \Psi_k\right) d^3 r}\\   	      
	       &=& {\int_P  \phi\left(\frac{\partial\rho}{\partial t}+\dvg\vec{j}\right)\,d^3r +
	          \int_V  \left( \vec{j}-\epsilon_0\epsilon_r \nonumber \frac{\partial\grad\phi}{\partial t} \right)
	          \sca\left(-\sum_{l=0}^N V_l \grad \Psi_k\right) d^3 r
	          - \int_P  \frac{\nu}{\epsilon_0\omega_{\rm pe}^2} \vec{j}^2 \, d^3r} \\  
	       &=&
	\sum_{l=0}^N V_l i_l 
  - \int_P  \frac{\nu}{\epsilon_0\omega_{\rm pe}^2} \vec{j}^2 \, d^3 r. \label{appinnerenergy}
\end{eqnarray}

\section{Properties of the dynamical operators}

In this section the properties of the dynamical operators are derived. It can be shown, that the operator of the conservative dynamics is anti-Hermitian
\begin{eqnarray}
\! \left\langle  z^\prime|T_C|z  \right\rangle \!+\! \left\langle  z^\prime\right|T_C^\dagger \left|z  \right\rangle \!\!&=&\!\! 
 \int_V \!\epsilon_0\epsilon_r \grad\phi\{\rho^\prime\}^*\sca\grad\phi\{-\dvg{\vec{j}}\}d^3r 
 + \int_P \!\frac{1}{\epsilon_0\omega_{\rm pe}^2}\vec{j}^{\prime*}\sca \left(-\epsilon_0 \omega_{\rm pe}^2 \grad \phi\{\rho\}\right)\!d^3r \nonumber\\
&&\!\!+ \int_V\!\epsilon_0\epsilon_r \grad\phi\{-\dvg{\vec{j}^\prime}\}^*\sca\grad\phi\{\rho\}d^3r 
 + \int_P\! \frac{1}{\epsilon_0\omega_{\rm pe}^2}\left(-\epsilon_0 \omega_{\rm pe}^2 \grad \phi\{\rho^\prime\}\right)^*\sca\vec{j}d^3r  \nonumber\\ 
 &=&\!\! 
 \int_V \!\phi\{\rho^\prime\}^* \dvg\left(\epsilon_0\epsilon_r \grad\phi\{\dvg{\vec{j}}\}\right) d^3r \nonumber
 + \int_P \! \dvg\vec{j}^{\prime*} \phi\{\rho\}\!d^3r \\
&&\!\!+ \int_V\!\dvg\left(\epsilon_0\epsilon_r \grad\phi\{\dvg{\vec{j}^\prime}\}^*\right)\phi\{\rho\}d^3r 
 + \int_P\!  \phi\{\rho^\prime\}^*\dvg \vec{j}d^3r  =0,\label{antihermitetc}
\end{eqnarray}
and the operator of dissipative dynamics is Hermitian
\begin{eqnarray}
 \left\langle  z'|T_D|z  \right\rangle - \left\langle  z'\right|T_D^\dagger \left|z  \right\rangle= 
 \int_P \!\frac{1}{\epsilon_0\omega_{\rm pe}^2}\vec{j}^{\prime*}\sca \left(-\nu\vec{j}\right)\!d^3r
 - \int_P \!\frac{1}{\epsilon_0\omega_{\rm pe}^2}\left(-\nu \vec{j}^{\prime*}\right)\sca \vec{j} \; d^3r
= 0\label{hermitetd}
\end{eqnarray}
Finally, we can show that it is negative definite
\begin{gather}
\left\langle  z|T_D|z  \right\rangle=
 \int_P \!\frac{1}{\epsilon_0\omega_{\rm pe}^2}\vec{j}^{*}\sca \left(-\nu\vec{j}\right)\!d^3r\leq 0.\label{posdeftd}
\end{gather}

\acknowledgments
The authors acknowledge the support by the Federal Ministry of Education and Research (BMBF) in frame of the project PluTO, and also the support by the Deutsche Forschungsgemeinschaft (DFG) via Graduiertenkolleg GK 1051, Collaborative Research Center TRR 87, and the Ruhr University Research School. Stimulating discussions with M. B\"oke, J. Winter and K. Nakamura are also acknowledged.

\pagebreak

\pagebreak

\pagebreak

\end{document}